\documentclass[fleqn,usenatbib]{mnras}

\usepackage{newtxtext,newtxmath}

\usepackage[T1]{fontenc}

\DeclareRobustCommand{\VAN}[3]{#2}
\let\VANthebibliography\thebibliography
\def\thebibliography{\DeclareRobustCommand{\VAN}[3]{##3}\VANthebibliography}

\usepackage{graphicx}	
\usepackage{amsmath}	
\usepackage{hyperref}
\usepackage{derivative}
\usepackage[dvipsnames]{xcolor}

\newcommand{\valpha}{\boldsymbol{\alpha}}
\newcommand{\inprept}[1]{#1 (\textit{in prep.})}
\newcommand{\inprepp}[1]{(#1 \textit{in prep.})}

\title[GPU Microlensing I]{Rootin' Tootin' Efficient Ray Shootin': Creating Microlensing Magnification Maps with GPUs}

\author[Weisenbach]{Luke Weisenbach$^{1,}$\thanks{E-mail: weisluke@alum.mit.edu}
\\
$^{1}$Institute of Cosmology and Gravitation, University of Portsmouth, Dennis Sciama Building, Burnaby Road, Portsmouth, PO1 3FX, UK
}

\date{Accepted XXX. Received YYY; in original form ZZZ}

\pubyear{2025}

\begin{document}
\label{firstpage}
\pagerange{\pageref{firstpage}--\pageref{lastpage}}
\maketitle

\begin{abstract}
The impending discovery and monitoring of hundreds of new gravitationally lensed quasars and supernovae from upcoming ground and space based large area surveys such as LSST, \textit{Euclid}, and \textit{Roman} necessitates the development of improved numerical methods for studying gravitational microlensing. We present in this work the fastest microlensing map generation code currently publicly available. We utilize graphics processing units to take advantage of the inherent parallelizable nature of creating magnification maps, in addition to using 1) the fast multipole method to reduce the runtime dependence on the number of microlenses and 2) inverse polygon mapping to reduce the number of rays required. The code is available at \url{https://github.com/weisluke/microlensing/}.
\end{abstract}

\begin{keywords}
gravitational lensing: micro -- methods: numerical -- gravitational lensing: strong
\end{keywords}



\section{Introduction}
\label{sec:intro}

Inverse ray shooting (IRS) has been the backbone of gravitational microlensing studies for the last nearly 40 years. Since its first usage by \cite{1986A&A...164..237S}, \cite{1986A&A...166...36K}, and \cite{1987A&A...171...49S}, ray tracing methods for microlensing have seen a multitude of developments and the appearance of competitive alternatives. First and foremost among the developments was an implementation of the \cite{1986Natur.324..446B} hierarchical tree code by \cite{1990PhDT.......180W, 1999JCoAM.109..353W}, which became the work horse of microlensing research for over a decade. A Fourier based method to calculate the long-range deflection of distant microlenses was later used by \cite{2004ApJ...605...58K} in order to generate magnification maps and fit large numbers of light curves for parameter estimation. The clever idea of inverse polygon mapping (IPM) was developed as an alternative by \cite{2006ApJ...653..942M}, achieving spectacular accuracy at lower computational costs. Parallel processing approaches on cluster Central Processing Units (CPUs) eventually came about \citep{2010NewA...15..181G}, soon to be rivalled by the appearance of Graphics Processing Unit (GPU) based methods \citep{2010NewA...15...16T} which matched the accuracy and speed of traditional IRS approaches \citep{2010NewA...15..726B} while having the advantage of free speedups from future hardware improvements due to Moore's law. This allowed for large parameter space studies such as GERLUMPH \citep{2013MNRAS.434..832V, 2014ApJS..211...16V} which had previously been computationally infeasible, if not impossible. 

Nearly a decade of relative quiet later, a host of improvements to these computational methods have further come about in the span of just a few years. This is in part due to a renewed theoretical and computational interest in microlensing, particularly in specific computationally difficult regions of parameter space, due to the discovery of individual stars highly magnified by galaxy clusters \citep{2017ApJ...850...49V, 2018NatAs...2..334K, 2022Natur.603..815W}. The anticipated explosion in the number of lensed quasars and supernovae from the Vera Rubin Observatory (LSST), \textit{Euclid}, and the \textit{Roman} Space Telescope in the upcoming decade has also motivated new investigations into numerical microlensing. Such improvements include a Poisson solver combined with IPM \citep{2021A&A...653A.121S} which reduces computation time, a clever scheme to reduce the number of microlenses necessary in simulations while taking advantage of GPU speedups \citep{2022ApJ...931..114Z}, and use of an extremely fast and accurate tree code for IPM \citep{2022ApJ...941...80J}. 

It is the latter of these improvements which inspired this work. Tree structures are ripe for implementation in microlensing codes on GPUs. While \texttt{teralens} \citep{teralens} has existed for at least 7 years and uses an implementation of the Barnes-Hut algorithm much like that of \cite{1990PhDT.......180W}, it has not appeared to gain much traction since its inception -- likely due to the unfortunate absence of any scientific publications. However, the fast multipole method (FMM) of \cite{1987JCoPh..73..325G} as suggested and used by \cite{2022ApJ...941...80J} on a CPU also has the potential to drastically decrease the computational costs of traditional IRS on GPUs while maintaining strict control on the tree code errors. Combined with an algorithm to apportion areas among pixels rather than relying solely on IRS, such a GPU version of an IPM code would be the fastest microlensing map generator currently known. The purpose of this paper\footnote{whose inception was driven primarily by the author's impatience while waiting on simulations to finish} is to present just such a code.

We implement the FMM and IPM to generate microlensing magnification maps on GPUs, achieving orders of magnitude speedups compared to both the direct GPU IRS methods \citep[e.g. GPU-D,][]{2010NewA...15...16T} and IPM on a CPU \citep{2022ApJ...941...80J}. While neither of these are surprising, given the hardware improvements since the creation of the GPU-D code and the unfair comparison between CPU and GPU codes, it is worth highlighting the drastic speedups possible when taking advantage of reasonably available modern computational resources. This work is, therefore, merely the latest in a long line of advances which we hope will be of use to the microlensing community. Our code is publicly available at \url{https://github.com/weisluke/microlensing/}; this paper serves to provide the necessary introduction to microlensing and magnification maps in Section~\ref{sec:background}, brief documentation of some of the computational ideas used for creating maps and reducing the runtime in Section~\ref{sec:comp_speedups}, and some benchmarks for the code in Section~\ref{sec:comp_timings}. As has been done in previous works, in Section~\ref{sec:extreme_mag_ex} we highlight the performance of our code using some specific sets of parameters of interest in order to demonstrate the current state of the art. We present our conclusions in Section~\ref{sec:conclusions}.

\section{Background}
\label{sec:background}

\subsection{(Micro)Lensing theory}

Gravitational lensing is a mapping from the image plane $\mathbf{x}$ to the source plane $\mathbf{y}$ via the projected two dimensional gravitational potential $\psi$ as \citep{1992grle.book.....S} \begin{equation}
    \mathbf{y} = \mathbf{x} - \valpha(\mathbf{x})
\end{equation} where the deflection angle $\valpha(\mathbf{x}) = \nabla\psi(\mathbf{x})$ and the equation has been non-dimensionalized for simplicity. If $\psi$ is taken to be the potential of the lens macromodel, microlensing adds a stochastic deflection angle $\valpha_\star(\mathbf{x})$ due to a field of point mass lenses (microlenses\footnote{typically assumed to be stars, but also any other sufficiently compact objects, e.g. planets, primordial black holes, or compact dark matter, which have Einstein radii much smaller than that of the macromodel, and whose physical extent is again much smaller than their Einstein radii}) and a corresponding deflection angle $\valpha_s$ from a sheet of smooth matter with equal but opposite mass density to compensate, so that the total mass density does not change. The lens equation then becomes \citep{1994A&A...288....1S} \begin{equation}
    \mathbf{y} = \mathbf{x} - \valpha(\mathbf{x}) - \valpha_\star(\mathbf{x}) - \valpha_s(\mathbf{x})
\end{equation} At a particular position in the image plane (e.g. the location of a macroimage), we can perform a Taylor expansion of the potential $\psi$. Centering our coordinate systems on the chosen image plane position and its corresponding source plane position, we can rewrite the lens equation as \begin{equation}
    \mathbf{y} = \mathbf{x} - \begin{pmatrix}
        \psi_{11} & \psi_{12}\\
        \psi_{21} & \psi_{22}
    \end{pmatrix}\mathbf{x} - \valpha_\star(\mathbf{x}) - \valpha_s(\mathbf{x})
\end{equation} where derivatives of the potential are evaluated at the chosen image plane position. This assumes that the length scale of microlensing is much smaller than the scale over which the macromodel potential varies and higher order derivatives of the potential are zero \citep[although, see e.g.][for the situation near a macromodel critical curve where this is no longer true; see also Appendix~\ref{app:macrocaustic}]{2017ApJ...850...49V}. Using the fact that the macromodel convergence is \begin{equation}
    \kappa = \frac{1}{2}(\psi_{11}+\psi_{22})
\end{equation} and the two components of the macromodel shear are \begin{equation}
    \begin{split}
    \gamma_1 &= \frac{1}{2}(\psi_{11}-\psi_{22})\\
    \gamma_2 &= \psi_{12}=\psi_{21}
    \end{split}
\end{equation} the lens equation becomes \begin{equation}
    \mathbf{y} = \mathbf{x} - \begin{pmatrix}
        \kappa + \gamma_1 & \gamma_2\\
        \gamma_2 & \kappa - \gamma_1
    \end{pmatrix}\mathbf{x} - \valpha_\star(\mathbf{x}) - \valpha_s(\mathbf{x})
\end{equation} We can additionally always rotate our coordinate system so that it aligns with the direction of the shear and use $\gamma = \sqrt{\gamma_1^2 + \gamma_2^2}$. We then have the standard microlensing lens equation \begin{equation}
    \mathbf{y} = \begin{pmatrix}
        1 - \kappa + \gamma & 0\\
        0 & 1 - \kappa - \gamma
    \end{pmatrix}\mathbf{x} - \valpha_\star(\mathbf{x}) - \valpha_s(\mathbf{x})
\end{equation} where the first term captures the deflection due to the macro-potential of the galaxy, and the latter two terms account for the stochastic deflection from the microlenses. 

The deflection angle from the microlenses is equal to \begin{equation}
    \valpha_\star(\mathbf{x}) = \theta_\star^2\sum_{i=1}^{N_\star}\frac{m_i(\mathbf{x}-\mathbf{x}_i)}{|\mathbf{x}-\mathbf{x}_i|^2}
\end{equation} where $m_i$ is the mass of a microlens, located at $\mathbf{x}_i$, in units of some mass $M$ (typically $M_\odot$) that determines the Einstein radius $\theta_\star$. The microlenses provide a convergence \begin{equation}
    \kappa_\star = \frac{\pi\theta_\star^2N_\star\langle m\rangle}{A_\star}
\end{equation} when distributed within some region of area $A_\star$. This convergence is related to the smooth matter fraction $s$ at the image location \begin{equation}
    s=1-\frac{\kappa_\star}{\kappa}
\end{equation} which is predominantly the dark matter fraction but
also includes any other smooth baryonic components.

\subsection{Complex lensing formalism}

Instead of vector quantities $\mathbf{x} = (x_1, x_2)$ and $\mathbf{y}=(y_1,y_2)$, one can use complex quantities \citep{1973ApJ...185..747B, 1975ApJ...195...13B, 1990A&A...236..311W} \begin{equation}
    \begin{split}
    z &= x_1 + ix_2\\
    w &= y_1 + iy_2
    \end{split}
\end{equation} to write the lens equation as \begin{equation}
    w = (1-\kappa)z + \gamma\overline{z} - \alpha_\star(z) - \alpha_s(z)
\end{equation} where \begin{equation}
    \alpha_\star(z) = \theta_\star^2\sum_{i=1}^{N_\star}\frac{m_i}{\overline{z} - \overline{z}_i}
\end{equation} This is a particularly useful programmatic choice which we adopt in our code. 

\subsection{Magnification maps}

\begin{figure}
    \centering
    \includegraphics[width=0.9\columnwidth]{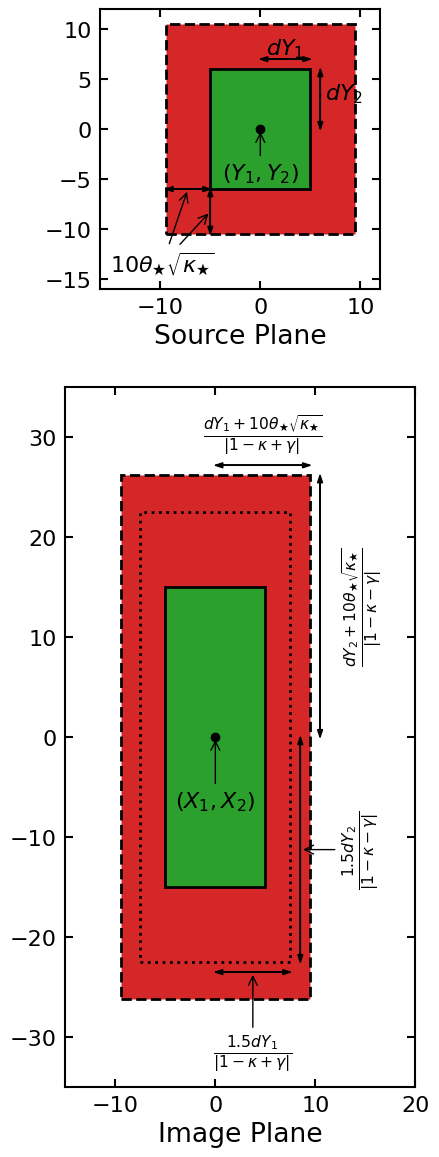}
    \caption{Visualization of the source and image plane regions under consideration when making a magnification map. The extent of the desired magnification map is the green solid border rectangle, and its image under just the macromodel is shown in the image plane. Due to stochastic deflections from the microlenses, a larger (red, dashed border rectangle) region in the image plane must be used for shooting rays. The size of the region comes from cross-correlating the desired source plane region with the PDF of the microlens deflection angle, which depends on $\kappa_\star$ and \textit{adds} to the required lengths. An arbitrary \textit{multiplicative} scaling factor (dotted image plane rectangle) is sometimes inadequate (see Figure~\ref{fig:small_map_comparison}).}
    \label{fig:macro_micro_areas}
\end{figure}

\begin{figure}
    \centering
    \includegraphics[width=\columnwidth]{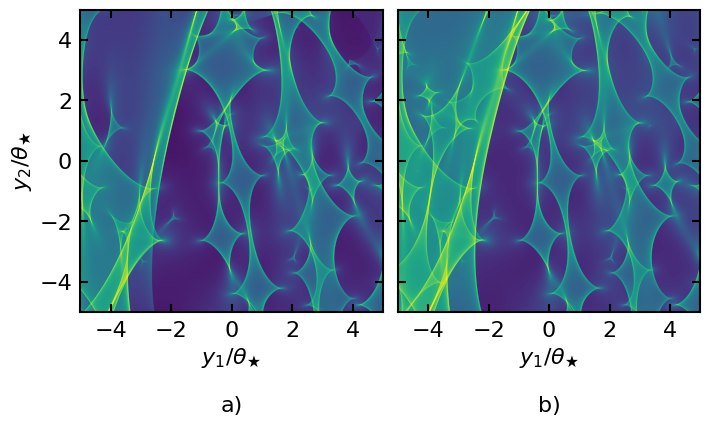}
    \caption{The magnification map on the left (a) was created using a shooting region 1.5 times larger than the source plane region mapped to the image plane under the macromodel. The map uses $(\kappa, \gamma, s)=(0.4,0.4,0)$ and a Salpeter mass spectrum for the microlenses. The same positions and masses of the microlenses were used to create the map on the right (b), but in that case the shooting region took into account $\kappa_\star$, $\langle m^2\rangle$, and $\langle m\rangle$ in determining its size. In map a), notice the absence of a few caustics going from the top left of the map towards the bottom left. Notice as well the incomplete small diamond caustic in the top right of map a) near the position $(4,3)$, along with spurious non-caustic lines. These come from not considering a large enough image plane region. This significantly affects some regions of the magnification map, and hence the magnification probability distribution and light curves.}
    \label{fig:small_map_comparison}
\end{figure}

By tracing photons (rays) from the image plane backwards to the source plane with the lens equation, one can build up a two dimensional array, i.e. map, of the number of rays that land at each source position. The source plane must be pixelated in order to accumulate rays, and the number of rays per pixel is proportional to the magnification of a source located at that pixel. 

If the microlensing map is a rectangular (arbitrarily, square typically) region centered at $(Y_1, Y_2)$ in the source plane with a corner located $(dY_1, dY_2)$\footnote{and assuming $dY_1 >0$, $dY_2 > 0$} away, then under the macromodel only the rays come from a rectangular image plane region with a center located at \begin{equation}
    (X_1, X_2) = \left(\frac{Y_1}{1-\kappa+\gamma}, \frac{Y_2}{1-\kappa-\gamma}\right)
\end{equation} and a corner that is \begin{equation}
    (dX_1, dX_2) = \left(\frac{dY_1}{|1-\kappa+\gamma|}, \frac{dY_2}{|1-\kappa-\gamma|}\right)
\end{equation} away. Due to the stochastic deflection of the microlenses however, a larger region in the image plane must actually be used when shooting rays. The size of this region comes from cross-correlating the desired source plane region with the probability distribution function (PDF) of the microlens deflection angle \textit{before} inverse mapping to the image plane with the macromodel. The cross-correlated source plane region has a corner that is farther away from the center, and the image plane region that needs considered is therefore a rectangle with a corner \begin{equation}
    (dX_1, dX_2) = \left(\frac{dY_1+ 10\theta_\star\sqrt{\kappa_\star}}{|1-\kappa+\gamma|}, \frac{dY_2+ 10\theta_\star\sqrt{\kappa_\star}}{|1-\kappa-\gamma|}\right)
\end{equation} away, which captures approximately 99\% of the flux from the microimages \citep{1986ApJ...306....2K}. See Figure~\ref{fig:macro_micro_areas} for a visualization; a slightly more detailed discussion of this is also given in Appendix~\ref{app:image_plane_area}. 

Some authors have previously used a multiplicative scaling factor on $(dY_1, dY_2)$ before inverse mapping with the macromodel in order to determine the size of the image plane region, e.g. \texttt{microlens}\footnote{\url{https://github.com/psaha/microlens}} \citep{1990PhDT.......180W}, the original IPM code of \cite{2006ApJ...653..942M}, \texttt{GPU-D}\footnote{\url{https://github.com/gvernard/GPU-D}} \citep{2010NewA...15...16T}, \texttt{mules}\footnote{\url{https://github.com/gdobler/mules}} \citep{2015ApJ...799..168D}, and \texttt{PIP}\footnote{\url{https://github.com/gilmerino/Microlensing-maps-generator}} \citep{2021A&A...653A.121S}; the scaling factor is an arbitrary choice which in some cases fails to capture the majority of the flux from the microimages. As \cite{2022ApJ...931..114Z} noted, there is an \textit{additive} border that depends on $\kappa_\star$ \textit{before} inverse mapping with the macromodel; see also \cite{1999MNRAS.306..223W} for a brief discussion of this in the context of creating microlensing light curves. 
While a multiplicative scaling factor may be appropriate, or more than appropriate, for some sets of parameters, it is not adequate for all -- especially when considering small source plane regions\footnote{The inadequacy of a multiplicative factor becomes obvious as one considers a source plane region that becomes pointlike; rays must clearly come from a non-pointlike region.}. Furthermore, for the case of a spectrum of masses, there is an additional dependence on $\langle m^2\rangle$ and $\langle m\rangle$ which must be taken into account \citep[][see also Appendix~\ref{app:image_plane_area}]{1986ApJ...306....2K} that can further drastically alter the size of the regions under consideration\footnote{This directly explains some of the computational difficulties encountered, e.g. in \cite{2020ApJ...904..176E}, when creating microlensing maps from a strongly bimodal mass distribution without accounting for the image plane region's dependence on $\kappa_\star$, $\langle m^2\rangle$, and $\langle m\rangle$.}. We showcase the importance of these considerations in Figure~\ref{fig:small_map_comparison}.

\section{Computational speedups}
\label{sec:comp_speedups}

In this section, we discuss the methods through which the computational time required to create magnification maps can be reduced.

\subsection{The number of microlenses}

The microlenses are distributed in a region somewhat larger than the required image plane region within which rays must be shot\footnote{It is here that a multiplicative factor is appropriate.} in order to avoid edge effects. The shape of the region for the microlenses is typically circular, as the form for $\valpha_s(\mathbf{x})$ then becomes very simple\footnote{Formally, $\alpha_s(\mathbf{x})$ should really be \begin{equation}
    \valpha_s(\mathbf{x}) = \begin{cases}
                -\kappa_\star\mathbf{x},& |\mathbf{x}| \leq R_\star\\
                \frac{-\kappa_\star\mathbf{x}R_\star^2}{|\mathbf{x}|^2},& |\mathbf{x}| > R_\star
            \end{cases}
\end{equation} where $R_\star$ is the radius of the circular region within which microlenses are distributed. Since the region of interest for IRS or IPM always lies within the microlens region however, it is only the first condition which is relevant.}: \begin{equation}
    \valpha_s(\mathbf{x}) = -\kappa_\star\mathbf{x}
\end{equation} or, using complex numbers\footnote{Formally again, $\alpha_s(z)$ should be \begin{equation}
    \valpha_s(z) = \begin{cases}
                -\kappa_\star z,& |z| \leq R_\star\\
                \frac{-\kappa_\star R_\star^2}{\overline{z}},& |z| > R_\star
            \end{cases}
\end{equation}}, $\alpha_s(z)=-\kappa_\star z$. 

However, for some systems the combination of $\kappa$ and $\gamma$ leads to a rectangular image plane region for shooting rays that has a large axis ratio and consequently a large number of microlenses if they are distributed in a circle. One can therefore instead distribute the microlenses in a rectangular region that is slightly larger than the rectangle of interest $(|X_1| + dX_1, |X_2| + dX_2)$; the form of $\valpha_s(\mathbf{x})$ is more complicated (see Appendix~\ref{app:alpha_smooth_rectangular} for the deflection angle of the smooth mass sheet $\alpha_s(z)$ if the microlenses are distributed in a rectangle), but the number of microlenses can be drastically decreased \citep{2022ApJ...931..114Z}. Some algebra shows that the decrease in the number of microlenses required when distributing them in a rectangular region is \begin{equation}
    \frac{N_{\star,\text{ rectangular}}}{N_{\star,\text{ circular}}}=\frac{2}{\pi(1 + 2\gamma^2|\mu_{\text{macro}}|)}
\end{equation} where the macromodel magnification \begin{equation}
    \mu_{\text{macro}}=\frac{1}{(1-\kappa)^2-\gamma^2}
\end{equation} For a moderately magnified macroimage with, e.g., $\kappa=\gamma=0.4$, distributing the microlenses in a rectangle requires approximately 4 times fewer than a circle.

The reduction in the number of microlenses from using a rectangular region can be important. Magnification maps created by the IRS method need on the order of hundreds or thousands of rays per pixel on average in order to reduce Poisson noise in the number of rays per pixel and achieve good statistics \citep{2013MNRAS.434..832V}. If the map is of a high resolution ($10^4$ pixels per side), this means that the number of rays required is of order $10^{10}-10^{11}$. Given that every microlens affects every ray, the runtime scales roughly $\propto N_{\text{rays}}N_\star$, and a factor of 10 reduction in the number of microlenses can reduce the calculation time by just as much. We will see however in Section \ref{sec:comp_timings} that, despite the reduction in the number of microlenses, the usage of a rectangular microlens region is not necessarily always a good idea.

\subsection{IPM and apportioning areas}

\begin{figure}
    \centering
    \includegraphics[width=\columnwidth]{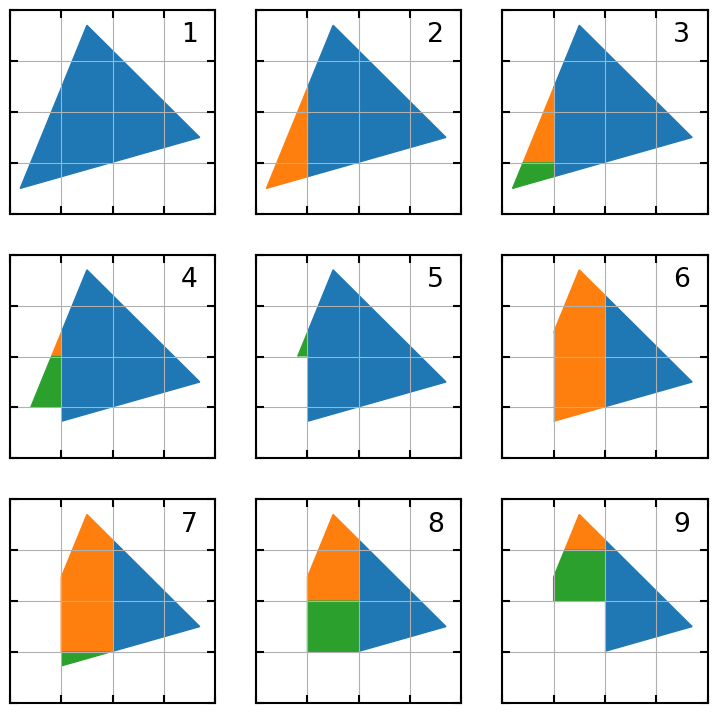}
    \caption{Visualization of the steps by which the area of a triangular cell is distributed among the pixels. The polygon (blue) is clipped into left (orange) and right (blue) polygons. The left polygon is then clipped into top (orange) and bottom (green) polygons. The bottom polygon has its area (as a fraction of the total cell area) added to its pixel, and the process repeats.}
    \label{fig:sutherland-hodgman}
\end{figure}

\begin{figure*}
    \centering
    \includegraphics[width=2\columnwidth]{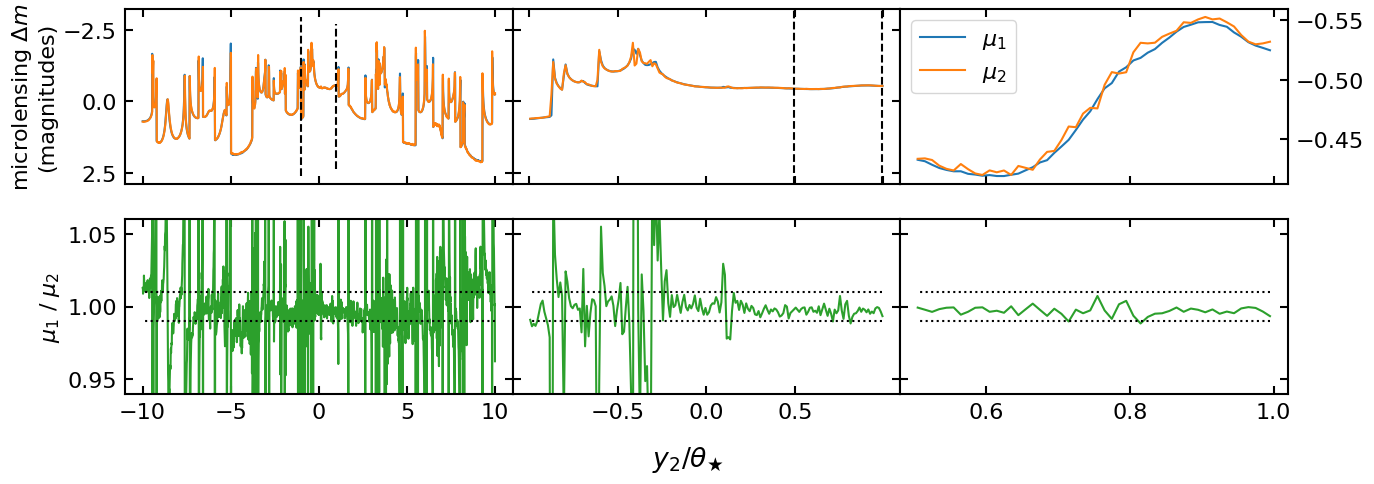}
    \caption{Light curves from maps created with our code ($\mu_1$) and the online tool \url{https://gloton.ugr.es/microlensing/} ($\mu_2$). There are minor differences due to the calculation of $\kappa_\star$ when reading in a file of microlens positions and masses, which slightly shifts the positions of the caustics creating the more noticeable deviations as sharp spikes. The light curves are essentially in agreement with each other though, to within a few percent (the dotted black lines in the bottom plots indicate the 1\% level), for the majority of the length.}
    \label{fig:lightcurve_comparisons}
\end{figure*}

\begin{figure}
    \centering
    \includegraphics[width=\columnwidth]{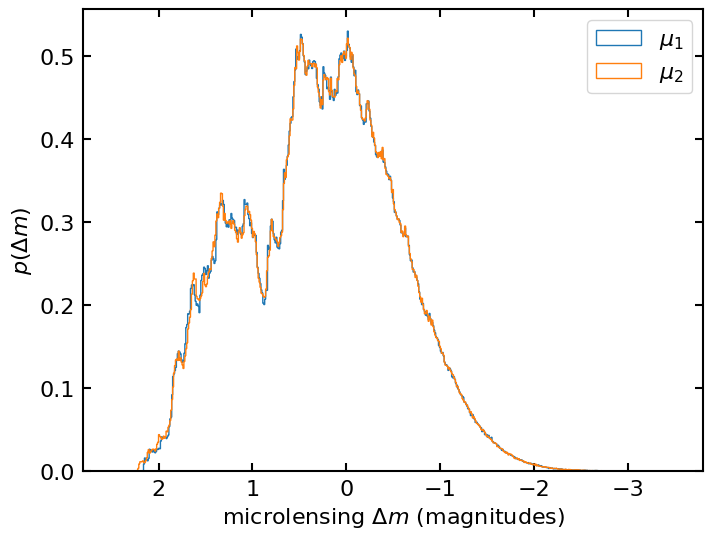}
    \caption{Magnification histograms for maps created with our code ($\mu_1$) and the online tool \url{https://gloton.ugr.es/microlensing/} ($\mu_2$), using the same microlens masses and positions. The distributions are essentially identical.}
    \label{fig:histogram_comparisons}
\end{figure}

IPM reduces the number of rays necessary by apportioning areas of regions of the image plane among the pixels of the source plane to which they are mapped, rather than accumulating rays within pixels like IRS, since magnification is a mapping of differential areas \begin{equation}
    \mu = \frac{\odif{A_{\text{image}}}}{\odif{A_{\text{source}}}}.
\end{equation} IPM can be used to achieve great accuracy over IRS when creating magnification maps; see \cite{2011ApJ...741...42M} for a thorough mathematical treatment of IPM. 

In brief, a rectangular (arbitrarily, square typically) grid of rays are mapped from the image plane to the source plane using the FMM; each rectangle defined by the grid of rays is referred to as a cell. We further split each rectangular cell into two triangular cells to avoid some of the issues encountered when a cell crosses critical curves \citep{2001astro.ph..2340K}. The areas of the cells once mapped to the source plane are then apportioned among the pixels of the source plane which they cover in order to create the magnification map. Compared to IRS, IPM can reduce the number of rays required per pixel to 1. While there are additional computational costs associated with apportioning areas as opposed to merely accumulating ray counts, IPM can still achieve 100-1000 times faster speeds on a CPU \citep{2006ApJ...653..942M}. Further improvements from a refined partitioning of the image plane can achieve the same accuracy while yet reducing computation time \citep{2011ApJ...741...42M}. 

We use the \cite{1974CACM...S} algorithm to apportion the areas. Each triangle is clipped along the columns and rows of the pixels which it intersects. Areas of the clipped regions are calculated with the shoelace formula (discrete version of Green's theorem in two dimensions), and as a fraction of the total area of the triangle are atomically incremented to the pixels. While we have not tested the Sutherland-Hodgman algorithm against the similar Sutherland-Cohen algorithm used by \cite{2021A&A...653A.121S}, we find our implementation adequate and straightforward enough to follow; see Figure \ref{fig:sutherland-hodgman}. 

We do not perform any subdivision of cells based on non-linearity conditions as in \cite{2006ApJ...653..942M, 2011ApJ...741...42M}, a decision also taken by \cite{2021A&A...653A.121S}. This is done as a compromise for accuracy and timing; \cite{2011ApJ...741...42M} note that a one-to-one correspondence between the size of the cells and the size of the pixels in the absence of lensing still leads to accurate magnification maps without such further refined partitioning.

\subsection{Tree codes}

\begin{figure}
    \centering
    \includegraphics[width=\columnwidth]{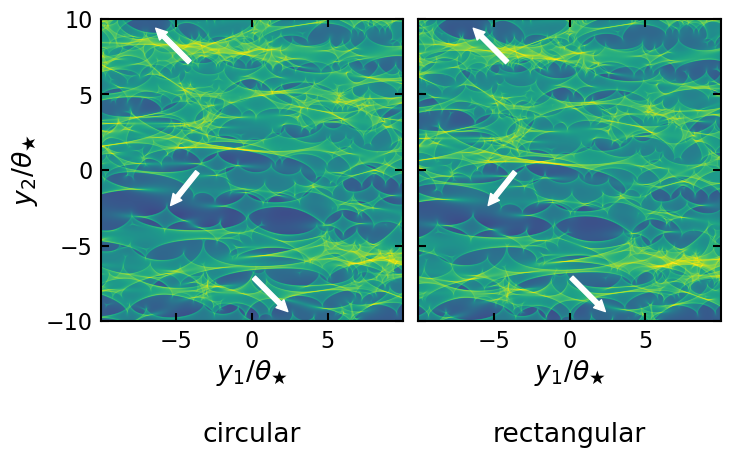}
    \caption{Two maps made with microlenses distributed in either a circular or rectangular region, where the positions of the microlenses in the rectangular region were reused for the circular region. While the maps display the same general features, the three white arrows mark regions where shifts in the locations of the caustics can be seen.}
    \label{fig:circ_rect_maps}
\end{figure}

\begin{figure}
    \centering
    \includegraphics[width=\columnwidth]{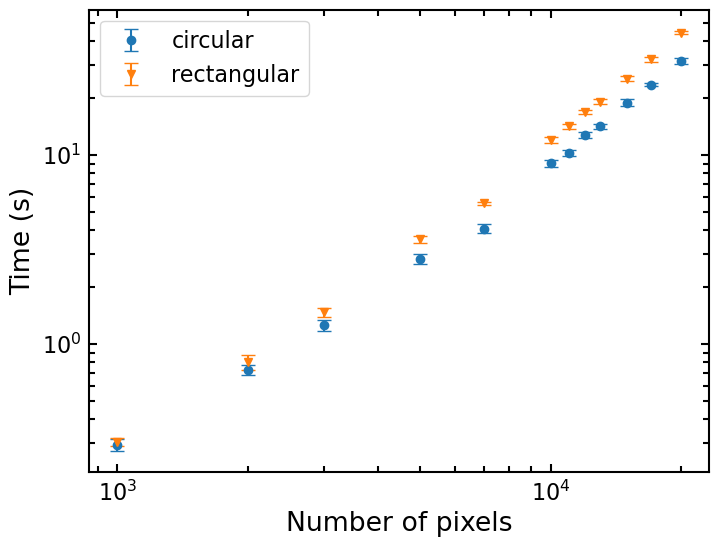}\\
    \includegraphics[width=\columnwidth]{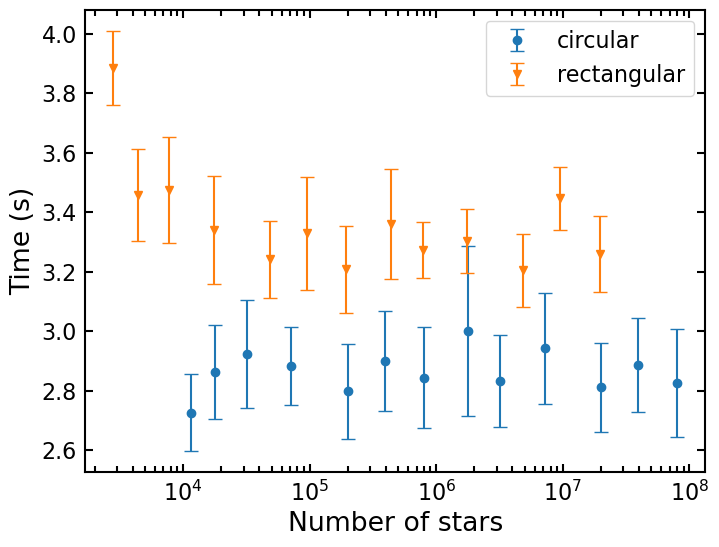}
    \caption{Timings (average and standard deviation of 10 simulations for each point) required to make magnification maps with IPM on an NVIDIA A100 80GB GPU for the parameters $\kappa=\gamma=0.4$, $\kappa_\star=0.2$. In all cases, the size of the magnification map was held fixed to $50\theta_\star$ x $50\theta_\star$, though the microlenses were distributed in either circular or rectangular regions. Top: The time required as a function of the number of pixels along each axis. The number of microlenses when distributed in a circle is $\sim15000$, while distributed in a rectangle is $\sim3600$. Bottom: The time required as a function of the number of microlenses. The number of pixels along each axis in this case was held fixed at 5000.}
    \label{fig:timing_analysis}
\end{figure}

Tree codes push all of the above improvements even farther by approximating the deflection angle due to distant microlenses, drastically reducing the number of microlenses that are used directly when shooting an individual ray. The \cite{1986Natur.324..446B} tree code achieves a computation time dependency that is $\propto N_{\text{rays}}\log N_\star$, while the FMM \citep{1987JCoPh..73..325G} has a dependency that is $\propto (N_{\text{rays}}+N_\star)$ -- a substantial improvement if $N_\star$ is less than the required $10^{8}-10^{11}$ rays. 

For the FMM, a tree structure is built starting with a root node which is a square that contains all the microlenses. If the number of microlenses in a given node and its neighbors\footnote{defined as any node which shares a corner or side with the given node} is more than the desired number of microlenses to use for directly calculating $\alpha_{\star}(z)$, the node is divided into four children; this process is continued until the desired maximum number of microlenses for  direct use is met\footnote{This is technically a slightly altered adaptive version from the non-adaptive method presented in \cite{1987JCoPh..73..325G}, or even the adaptive method presented in \cite{1988SJSSC...C}, as we consider the number of microlenses in a node and its neighbors; see Appendix~\ref{app:fmm}.}. Inside a given node, the deflection angle due to the microlenses is broken into two components: one from nearby microlenses (i.e. contained within the node and its neighbors), and one from far away microlenses, as \begin{equation}
    \alpha_\star(z) = \alpha_{\star,\text{ near}}(z) + \alpha_{\star,\text{ far}}(z)
\end{equation} The deflection angle from nearby microlenses is calculated directly. The deflection angle from far away microlenses $\alpha_{\star,\text{ far}}(z)$ is equal to the conjugate of the derivative of the potential from the distant microlenses \citep{1992grle.book.....S}, \begin{equation}
    \alpha_{\star,\text{ far}}(z) = \overline{\frac{\partial \psi_{\star,\text{ far}}(z)}{\partial z}}
\end{equation} The potential $\psi_{\star,\text{ far}}(z)$ is locally approximated within a node by a Taylor series, which itself comes from approximating and summing the multipole expansions of distant microlenses and nodes; the multipole and Taylor coefficients are straightforwardly calculated from the equations in \cite{1987JCoPh..73..325G} when using complex coordinates since the gravitational potential due to point mass lenses is equivalent to the electrostatic potential from point charges as considered in that work. 

We implement the FMM ourselves rather than requiring an external library. Some minor details of the implementation are given in Appendix~\ref{app:fmm}. In essence, we end up with a collection of nodes which each contain \begin{itemize}
    \item[1.] the microlenses to directly use, and
    \item[2.] coefficients of a Taylor series which locally approximates within the node the deflection angle from distant microlenses.
\end{itemize} We set the maximum number of microlenses allowed to be directly used to 32. The order of the Taylor series depends on the size of the map, the pixel scale, and the desired accuracy when shooting rays\footnote{which we set to 1/10 of the smallest side length of a pixel}, but is typically less than 30\footnote{And indeed we set the maximum allowable order to 31, for memory purposes -- i.e. so we have at most 32 terms including the zeroth order.}. Tracing a ray from the image plane to the source plane is thus reduced from a sum that depends on all of the microlenses (anywhere from hundreds to millions), to a sum that depends on at most 32 microlenses and a polynomial with $\sim$30 terms -- quite a substantial reduction!

\subsection{GPUs}

The base process of tracing rays is embarrassingly parallel as every light ray is independent of the others, and GPUs can therefore be used to speed up codes by even greater amounts \citep{2010NewA...15...16T, 2010NewA...15..726B, 2014A&C.....6....1V, teralens}. We implement IPM and the FMM with NVIDIA's CUDA, introducing orders of magnitude improvements which we highlight in Section~\ref{sec:comp_timings}.

\section{Comparisons and Timing}
\label{sec:comp_timings}

We perform two tests of our code: 1) a comparison of magnifications and 2) a timing analysis to demonstrate how the runtime scales with various parameters. 

\subsection{Magnification comparisons}

We perform two different comparisons of magnifications. For the first, we download a selection of maps and the microlenses which created them from the online tool at \url{https://gloton.ugr.es/microlensing/} created by \cite{2022ApJ...941...80J}\footnote{While \cite{2021A&A...653A.121S} also provide an online tool to create magnification maps at \url{https://microlensing.overfitting.es}, it does not appear to offer the option to download the information of the microlenses used.}. Figure~\ref{fig:histogram_comparisons} shows the magnification distributions for a particular set of parameters, $\kappa=\kappa_\star=\gamma=0.6$. The histograms are essentially indistinguishable. Cases for other sets of parameters examined are similarly indistinguishable, and we do not show them here. Light curves, shown in Figure~\ref{fig:lightcurve_comparisons}, are nearly indistinguishable as well, in agreement to within $\pm1$\% for the majority of the light curve length. Differences are more pronounced near the edges where the importance of the image plane region used to shoot the cells comes into play.

For the second comparison of magnifications, we perform a consistency check on our code. We create magnification maps for a set of parameters using IPM and IRS with and without using the FMM. We additionally use both rectangular and circular microlens regions, where the positions of the microlenses from the rectangular region are reused in the circular one so that we can check whether the magnification map drastically changes or not -- a comparison which was not performed in \cite{2022ApJ...931..114Z}. We do not find any differences between maps made with or without the FMM, nor any differences between maps made from IPM or IRS. However, despite sharing microlenses with the same positions over a large region, the maps made from circular and rectangular regions of microlenses are slightly different as shown in Figure~\ref{fig:circ_rect_maps}. While we find that the magnification distributions are in agreement with each other, a full study of whether there are significant alterations to the magnification distributions elsewhere in parameter space or to lightcurves is outside the scope of this work. There are no theoretical reasons however to believe the shape of the microlens region should drastically affect any statistics so long as the average deflection of the microlenses is consistently accounted for. 

\subsection{Timing analysis}

We show in Figure~\ref{fig:timing_analysis} the time required to make magnification maps as a function of two parameters: the number of pixels and the number of microlenses. We might expect the runtime to scale quadratically with the number of pixels (taken to be the same along each axis), and find this to be roughly true -- though we note the relation is not exactly quadratic, as increasing the number of pixels by a factor of 10 does not quite increase the runtime by a factor of 100. The runtime dependence on the number of microlenses is found to be essentially non-existent up to a large number ($N_\star\sim10^8$) of microlenses -- consistent with the expectation from the FMM that it is the larger number of rays which dominates the runtime. While we do not use the same microlensing parameters as, e.g. Figure 7 of \cite{2022ApJ...941...80J}, comparing their figure with the times shown in Figure~\ref{fig:timing_analysis} highlights that we can achieve a speedup of a factor of roughly 100 or more in magnification map generation by utilizing GPUs.

We note here that distributing the microlenses in a rectangular region is not necessarily useful for creating magnification maps as the FMM essentially removes any runtime dependency on $N_\star$. The more complicated expression for $\alpha_s(z)$ when the microlenses are distributed in a rectangle increases the runtime due to the presence of multiple logarithmic terms, as opposed to a single multiplication for $\alpha_s(z)$ when the microlenses are distributed in a circle. Distributing the microlenses in a rectangle is, however, useful for other microlensing purposes such as finding the microlensing critical curves and caustics \inprepp{Weisenbach}, and so therefore still has merit.

We show in Table~\ref{tab:runtimes} the time taken to create the magnification maps considered for the magnification comparisons, which makes more visible the speedups available on a GPU from the FMM, IPM, and using a rectangular or circular microlens region, when compared to IRS with no improvements. Using IPM as opposed to IRS is a substantial gain, as is reducing the number of microlenses directly used. However, the FMM is by far the biggest factor in improving the runtime. With the FMM, one can see how the complexity of terms in the smooth deflection angle for a rectangular region of microlenses increase the time taken. IPM is also slightly slower on a GPU compared to IRS due to the time required for apportioning areas -- though given that IPM produces less noisy maps, the tradeoff is much more acceptable.

\begin{table}
 \caption{The time taken (in seconds) to create a magnification map on an NVIDIA A100 80 GB GPU under various considerations. The map was kept fixed at $20\theta_\star$ x $20\theta_\star$, 2000 x 2000 pixels. The number of microlenses was $\sim$27,000 (3000) for the circular (rectangular) region. The number of rays per pixel in the absence of lensing was 1 for IPM and 100 for IRS.}
 \label{tab:runtimes}
     \begin{center}
         \begin{tabular}{ccc}
          \hline
           & IPM & IRS\\
          \hline
          no FMM & & \\
          \hline
          circular & 39.367 & 348.884\\
          rectangular & 4.999 & 37.663\\
          \hline
          FMM & & \\
          \hline
          circular & 0.763 & 0.748\\
          rectangular & 0.839 & 2.141\\
          \hline
         \end{tabular}
     \end{center}
\end{table}

\section{Extreme magnification examples}
\label{sec:extreme_mag_ex}

\begin{figure*}
    \centering
    \includegraphics[width=2\columnwidth]{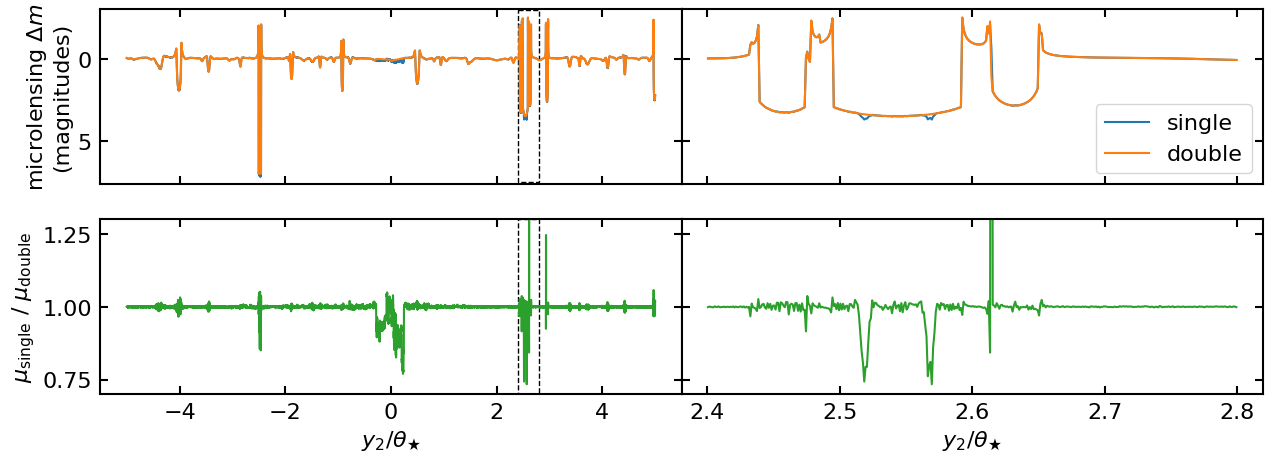}
    \caption{Top: Light curves for a vertical slice of the maps in Figure \ref{fig:single_double_maps} where $y_1=-21$. Bottom: Ratio of the single precision magnification to the double precision magnification. The right panels are zooms of the dashed rectangular regions indicated in the left panels. While the single-to-double precision magnifications are generally scattered around 1, there are some regions where floating point precision loss leads to substantial ($\sim$20\%) underestimates of the magnification. We note that the large spikes caused by differences near caustic crossings are to be expected due to the extreme magnification changes over $\sim 1$ pixel scales.}
    \label{fig:single_double_lightcurves}
\end{figure*}

\begin{figure}
    \centering
    \includegraphics[width=\columnwidth]{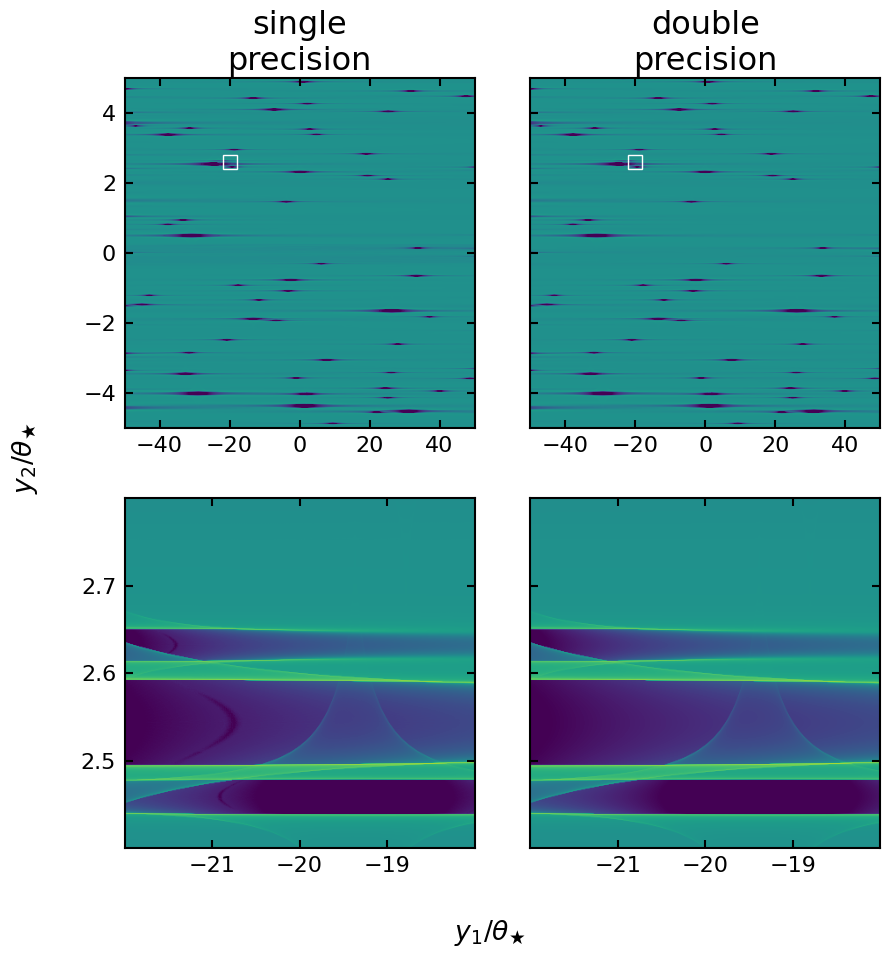}
    \caption{Top: Magnification maps created using single precision (left) and double precision (right). Bottom: Zoomed in regions for the white squares indicated in the top panels. While on a large scale there appears to be no differences between the maps, there are residual features on small scales, e.g. around the point (-20.75, 2.55), due to floating point precision loss when mapping a large image plane region. This affects the light curves as shown in Figure \ref{fig:single_double_lightcurves}.}
    \label{fig:single_double_maps}
\end{figure}

\begin{figure}
    \centering
    \includegraphics[width=\columnwidth]{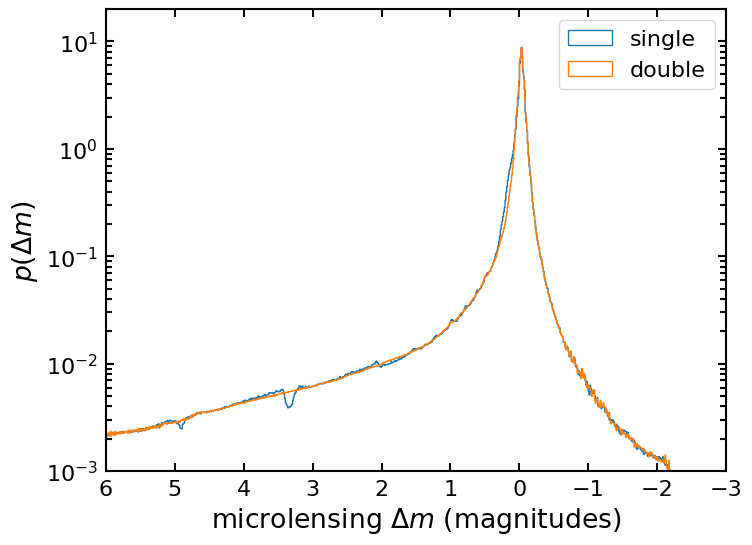}
    \caption{Histogram of the microlensing (de)magnification for the single and double precision maps of Figure~\ref{fig:single_double_maps}. While individual light curves taken from each map will differ, potentially substantially as shown in Figure~\ref{fig:single_double_lightcurves}, the magnification histograms are largely in agreement with each other.}
    \label{fig:single_double_hists}
\end{figure}

\begin{figure*}
    \centering
    \includegraphics[width=2\columnwidth]{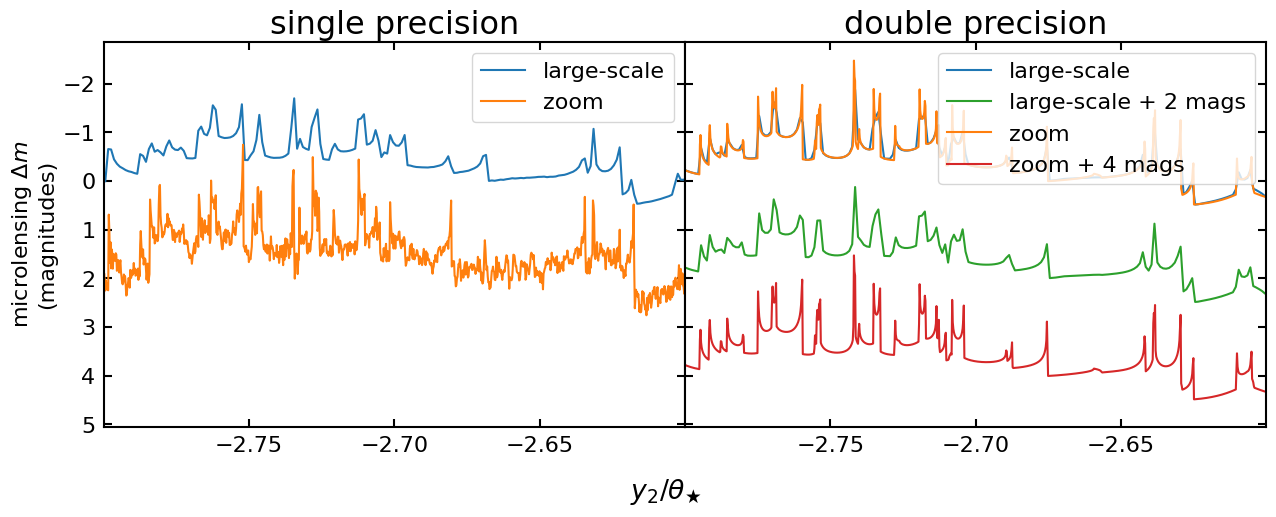}
    \caption{Left: Light curves created from the large scale single precision map of Figure~\ref{fig:low_s_maps} as well as the zoomed higher resolution map. Right: Light curves from the large scale double precision map and the higher resolution zoom. As seen in the left, floating point precision loss can lead to a substantial (2 magnitude) difference in the light curves when one recreates a portion of the map at higher resolution. Furthermore, the many peaks seen in the light curve are not caustic crossings as one might expect, but rather just noise from precision loss. The double precision light curves for both the large scale map and the higher resolution zoom agree on the magnifications. We further shift the two double precision light curves so they are separately visible, illustrating that the zoomed version is indeed at higher resolution, resolving some of the caustic crossings while maintaining the smoothness of the light curve. We note that while the light curves on the left and right are meant to show the same cut through the source plane, there are minor positional shifts due to slight differences in single/double precision calculations of $\kappa_\star$ when reading in the file of microlenses.}
    \label{fig:low_s_lightcurves}
\end{figure*}

\begin{figure}
    \centering
    \includegraphics[width=\columnwidth]{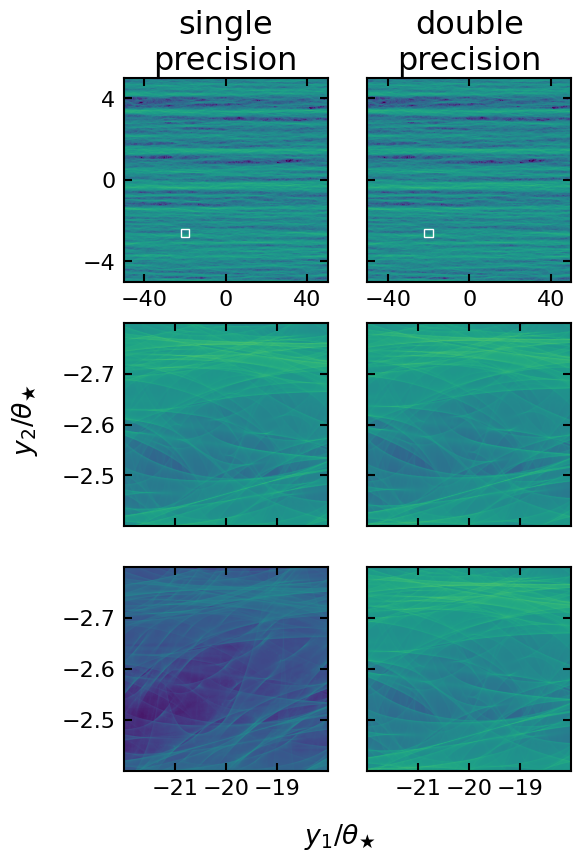}
    \caption{Top: Magnification maps created using single precision (left) and double precision (right) for higher stellar mass density than Figure~\ref{fig:single_double_maps}. Middle: Zoom of the large maps for the white squares indicated in the top panels. Bottom: Maps of the indicated white squares recreated with 5 times the resolution. The smaller physical scale of the pixels can exacerbate the errors induced by floating point precision when calculating magnifications for the single precision map.}
    \label{fig:low_s_maps}
\end{figure}

As mentioned in the introduction, of particular interest recently in microlensing research is the regime of high magnification ($\mu \gtrsim 1000$), typically near the critical curves of galaxy clusters. The interested readers are referred to \cite{2017ApJ...850...49V}, \cite{2018PhRvD..97b3518O}, \cite{2018ApJ...857...25D}, \cite{2018NatAs...2..334K}, \cite{2022Natur.603..815W}, references therein, and citations thereof. While some recent research has focused on refined methods for creating individual microlensing light curves \citep{2022A&A...665A.127D, 2022MNRAS.514.2545M}, the need for magnification maps in order to create many light curves at once and to analyze magnification statistics still plays a vital role \citep{2024A&A...687A..81P}. We take for comparison here a set of parameters used in Appendix B of \cite{2024A&A...687A..81P}, \begin{equation}
    \begin{split}
        (1-\kappa+\gamma)^{-1}&= 1.5,\\
        (1-\kappa-\gamma)^{-1}&= -1000,\\
        s&= 1-5\cdot10^{-5}
    \end{split}
\end{equation} which presumably took $\sim$30 CPU-hours to compute (the stated average for their simulations) for code parallelized to run on cluster CPUs. At $z_{\text{lens}}=0.7$ and $z_{\text{source}}=1.3$ (as used in that work), the Einstein radius of a $M_\odot$ star is $\approx2$ micro-arcseconds. We consider a source plane of size 100 by 10 Einstein radii, and 10,000 by 10,000 pixels. This choice of different physical scales along each pixel axis, while different from \cite{2024A&A...687A..81P}, can be justified by the preferential compression direction of the macromodel, along which we want high resolution to resolve the microcaustics. We use microlenses with a Salpeter mass spectrum distributed between $0.1M_\odot$ and $2M_\odot$.

Figure~\ref{fig:single_double_maps} shows an example magnification map for this set of parameters. We create the map using both single and double precision, which took $\sim$200 and $\sim$600 seconds respectively\footnote{Though we would expect the double precision case to take only twice the amount of time as single precision since the GPU reports a performance ratio of 2 between them, this does not quite appear to be the case.}. While single precision is likely adequate on large scales for looking at magnification statistics (e.g. Figure~\ref{fig:single_double_hists})\footnote{We note the absence of any substructure around the peak of the distribution in Figure~\ref{fig:single_double_hists}, compared to Figure B1 of \cite{2024A&A...687A..81P}. While we do not have any conclusive reasons for this, the likely suspect for the difference in that work is improperly accounting for the appropriate size of the source plane from the considered image plane, stellar density, and stellar mass spectrum.}, floating point precision loss can lead to artifacts in light curves which may be undesirable (e.g. Figure~\ref{fig:single_double_lightcurves}). 

A more extreme version of this can be illustrated by increasing the surface mass density in microlenses to $\kappa_\star = 0.01\kappa$ ($s=0.99$), closer to the values expected in galaxy clusters from the intracluster medium. We create maps of the same size and pixel scale as before\footnote{which took $\sim$700 and $\sim$2400 seconds for single and double precision, respectively, due to the substantial increase in $\kappa_\star$. This set of parameters produced a number of microlenses and cells comparable to that considered in Section 3.3 of \cite{2022ApJ...941...80J}, which took $\sim$1 day to simulate on a CPU.}. We then consider zooming in on a particular region with 5 times the resolution (i.e. creating a map centered at some new location, using the same microlenses and number of pixels, but with a size of 20 by 2 Einstein radii). Due to the smaller physical scale of the pixels, floating point errors are more extreme at higher resolution as evident in the bottom left panel of Figure~\ref{fig:low_s_maps}. This can also be seen in the light curves, as shown in Figure~\ref{fig:low_s_lightcurves}. Unsurprisingly, when simulations must take into consideration extremely large image plane regions, one needs to take into account computer precision as well -- which can drastically alter the runtime and results of simulations.

\section{Conclusions}
\label{sec:conclusions}

We have presented the current state of the art in microlensing map generation. The FMM efficiently approximates the deflection angles of distant microlenses \citep{1987JCoPh..73..325G}, while IPM reduces the number of rays required per pixel \citep{2006ApJ...653..942M}. We implement both on GPUs to take advantage of the inherent parallelizable nature of microlensing. There are no remaining improvements that could significantly reduce the computational runtime required outside of altering the code to run on multiple GPUs. 

The code is flexible, able to cover the entirety of microlensing parameter space, consistent with known theoretical requirements for capturing the majority of the microimage flux within a given pixel \citep{1986ApJ...306....2K}, able to handle generic map sizes (both physical and pixel), and capable of handling a variety of microlens mass distributions as well as spatial distributions including clustered or uniform random. 

Our code is applicable not only for current microlensing research such as creating maps to be used in the modeling of lensed quasars, supernovae, and individual high redshift stars near galaxy cluster caustics, but it will also prove useful in new areas of research. Dynamical models that account for the motion of the microlenses and studies of the impact of the microlens mass spectrum in particular regions of parameter space are now much more computationally feasible to simulate.

We welcome any collaborations, and encourage interested users to both use the code and reach out with any questions, bugs, or improvements.

\section*{Acknowledgements}

The author would like to thank Dan Ballard for\footnote{somewhat facetiously} providing the title of this paper. He would also like to thank James Chan, Giorgos Vernardos, Timo Anguita, Arjun Murlidhar, Sai Vidyud Senthil Nathan, Sinclaire Jones, Scott Gaudi, Paras Sharma, and Padma Venkatraman for useful discussion, testing, and input. He would like to thank Tom Collett as well for his continued support throughout the author's PhD. Lastly\footnote{and against his better wishes}, the author would also like to thank Dan Ryczanowski for suggesting the\footnote{abhorrent} acronym \texttt{RooTERS}, which he has had to respectfully decline using\footnote{Users of the code are under no such obligation to avoid this acronym -- though the author would prefer if they do.}.

Numerical computations were done on the Sciama High Performance Compute (HPC) cluster which is supported by the ICG, SEPNet, and the University of Portsmouth. 

This work has received funding from the European Research Council (ERC) under the European Union’s Horizon 2020 research and innovation programme (LensEra: grant agreement No. 945536). For the purpose of open access, the authors have applied a Creative Commons Attribution (CC BY) license to any Author Accepted Manuscript version arising.

\section*{Data Availability}

Data from this work can be made available upon reasonable request to the corresponding author. The code developed is publicly available and linked in the abstract and introduction. Bugs can be raised as issues on github or reported to the author via email. Running the code requires an NVIDIA graphics card. The NVIDIA CUDA compiler \texttt{nvcc} is required to compile the code, as well as a C++20 compliant compiler. Precompiled libraries created using the GNU compiler v11.2.0 and the CUDA compiler v12.4 are also provided, which should work on Linux distributions that have GLIBC >= 2.31 and GLIBCXX >= 3.4.29.



\bibliographystyle{mnras}
\bibliography{bibliography.bib} 




\appendix

\section{How large of an image plane area to use}
\label{app:image_plane_area}

For convenience, we summarize here arguments from other works which highlight the required image plane region to be considered in microlensing simulations.

The magnification of a point source located at $\mathbf{y}$ can be written as \citep{2003A&A...404...83N, 2017ApJ...850...49V} \begin{equation}
    \mu(\mathbf{y}) = \int \delta(\mathbf{x} - \valpha(\mathbf{x}) - \valpha_\star(\mathbf{x}) - \valpha_s(\mathbf{x}) - \mathbf{y})\odif[order=2]{\mathbf{x}}
\end{equation} The combination of the terms $\valpha_\star$ and $\valpha_s$ can be viewed as a random variable $\valpha' = \valpha_\star + \valpha_s$ that changes based on different realizations of the random point mass positions. Averaging over all such realizations\footnote{and ignoring the formal mathematics covered in detail by others, which typically requires moving to Fourier space} \citep{2017ApJ...850...49V, 2021arXiv210412009D}, \begin{equation}
    \langle\mu(\mathbf{y})\rangle = \int \delta(\mathbf{x} - \valpha(\mathbf{x}) - \valpha' - \mathbf{y})p(\valpha')\odif[order=2]{\valpha'}\odif[order=2]{\mathbf{x}}
\end{equation} where $p(\valpha')$ is the PDF of the deflection angle from the microlenses \citep{1986ApJ...306....2K, 1992grle.book.....S, 2009JMP....50g2503P}.

By transforming coordinates from the image plane to the source plane with the macromodel using the change of variable \begin{equation}
     \mathbf{y}' = \mathbf{x}- \valpha(\mathbf{x}) - \mathbf{y}
\end{equation} \cite{2017ApJ...850...49V} showed that this simplifies to \begin{equation}
    \langle\mu(\mathbf{y})\rangle = \int p(\mathbf{y}') \mu_{\text{macro}}(\mathbf{y}' + \mathbf{y}) \odif[order=2]{\mathbf{y}'}.
\end{equation} 
This expression is a cross-correlation of the microlens deflection PDF with the magnification of the macro model\footnote{Technically contrary to what is often stated; it is not a convolution since the kernel is not reversed. Since the deflection angle PDF is radially symmetric however, there is no difference.}. What was once a point source has, on average, been ``smeared out'' into a source with a profile that looks like $p(\valpha')$. This argument was laid out originally by \cite{1986ApJ...306....2K} in different notation, and also presented in \cite{1992grle.book.....S}, but we found the presentation by \cite{2017ApJ...850...49V} and \cite{2021arXiv210412009D} particularly enlightening enough to include the above. 

For an extended source, a similar process can be done with the conclusion that the average magnification is that of the source profile cross-correlated with the PDF of the microlens deflection and subject to the macromodel magnification \citep{2021arXiv210412009D}. 

In the context of microlensing maps, our extended source profile is a rectangle. The microlens deflection angle PDF is isotropic, and approximately 99\% of the PDF is contained within a radius of $r=10\theta_\star\sqrt{\kappa_\star}$ \citep{1986ApJ...306....2K}. The cross-correlation of the rectangular source plane region with the microlens deflection angle PDF can be approximated as simply adding a border of width $r$ around the entirety of the rectangular region. Transforming this new rectangular region to the image plane using the macromodel gives the required region in the image plane which must be considered.

More accurately, the tail of the microlensing PDF behaves like \citep{1986ApJ...306....2K} \begin{equation}
    p(|\valpha'|)=\frac{\theta_\star^2\kappa_\star}{\pi|\valpha'|^4}\frac{\langle m^2\rangle}{\langle m\rangle}
\end{equation} when accounting for the mass spectrum of the microlenses as well; \cite{1991A&A...250...62R} also derive a microlensing dependence on $\langle m^2\rangle / \langle m\rangle$, though in a slightly different context. This means that on average a fraction $f$ of the PDF lies outside the radius \begin{equation}
    r = \theta_\star\sqrt{\frac{\kappa_\star\langle m^2\rangle}{f\langle m\rangle}}
\end{equation} so long as $f \lesssim 1 / 100$. We note that $f = 1 / 1000$ is a more conservative, but better \citep[in light of, e.g., Figure 2 of][]{1986ApJ...306....2K} approximation which we adopt as default.

\section{Complex deflection angle for a rectangular mass sheet}
\label{app:alpha_smooth_rectangular}

The deflection angle for a rectangular mass sheet in vector coordinates is given in \cite{2022ApJ...931..114Z}. Using complex notation, the deflection angle can be found from \citep{1973ApJ...185..747B, 1975ApJ...195...13B, 1992grle.book.....S}

\begin{equation}
    \overline{\alpha(z)} = \frac{1}{\pi}\int\kappa(z')\frac{1}{z-z'}\odif[order=2]{z'}
\end{equation} For a sheet of convergence $-\kappa_\star$ centered at $(0,0)$ with a corner at $(c_1, c_2)$, $c_1 > 0$, $c_2 > 0$, the deflection angle $\alpha_s(z)$ is \begin{equation}
    \overline{\alpha_s(z)} = \frac{-\kappa_\star}{\pi}\int_{-c_2}^{c_2}\int_{-c_1}^{c_1}\frac{1}{z - (x_1' + ix_2')}\odif{x_1'}\odif{x_2'}
\end{equation} Using $u=z - (x_1'+ix_2')$, $\odif{u} = -\odif{x_1'}$ we have

\begin{equation}
    \begin{split}
    \overline{\alpha_s(z)} &= \frac{-\kappa_\star}{\pi}\int_{-c_2}^{c_2}\int_{z-(c_1+ix_2')}^{z-(-c_1+ix_2')}\frac{1}{u}\odif{u}\odif{x_2'}\\
    &= \frac{-\kappa_\star}{\pi}\int_{-c_2}^{c_2} \log[z-(-c_1+ix_2')] - \log[z-(c_1+ix_2')]\odif{x_2'}
    \end{split}
\end{equation} Using $v = z - (\pm c_1 + ix_2')$, $\odif{v} = -i\odif{x_2'}$, we have 

\begin{equation}
    \overline{\alpha_s(z)} = \frac{-\kappa_\star i}{\pi}\left[\int_{z-(-c_1-ic_2)}^{z-(-c_1+ic_2)} \log v \odif{v} - \int_{z-(c_1-ic_2)}^{z-(c_1+ic_2)}\log v \odif{v}\right]
\end{equation} Denoting the corner of the mass sheet as $c = c_1+ic_2$, we simplify

\begin{equation}
    \begin{split}
    \overline{\alpha_s(z)} = \frac{-\kappa_\star i}{\pi}\Big[&\int_{z+c}^{z+\overline{c}}\log v \odif{v} - \int_{z-\overline{c}}^{z-c}\log v \odif{v}\Big]\\
    = \frac{-\kappa_\star i}{\pi}\Big[&(z+\overline{c})\log(z+\overline{c})-(z+\overline{c})\\
    & - (z+c)\log(z+c) + (z+c)\\
    & -(z-c)\log(z-c)+(z-c)\\
    & +(z-\overline{c})\log(z-\overline{c})-(z-\overline{c})\Big]\\
    = \frac{-\kappa_\star i}{\pi}\Big[&(z+\overline{c})\log(z+\overline{c}) - (z+c)\log(z+c)\\
    & -(z-c)\log(z-c) +(z-\overline{c})\log(z-\overline{c})\Big]\\
    =\frac{-\kappa_\star i}{\pi}\Big[&(c-z)\log(c-z) - (\overline{c}-z)\log(\overline{c}-z)\\
    & +(-c-z)\log(-c-z)-(-\overline{c}-z)\log(-\overline{c}-z)\Big]
    \end{split}
\end{equation} where in the last step, we reorder some of the terms and flip the sign on all the combinations of $z$ and $c$ that appear so that they are more readily seen as representations of the complex distance from $z$ to one of the corners of the rectangle. Symmetries ultimately cancel the additional factors that appear from flipping the signs inside the logarithms\footnote{One could also just factor a negative sign from the denominator in the beginning of the calculations to arrive at the same conclusions.}.

However, we must take care with branch cuts of the natural logarithm. We take the branch cut as usual to be the negative real axis. When integrating over $\odif{x_2}$, we cross the branch cut between $-c-z$ and $-\overline{c}-z$ if $\operatorname{Re}(z) \geq -c_1$; we cross the branch cut as well between $\overline{c}-z$ and $c-z$ if $\operatorname{Re}(z) \geq c_1$. In addition, these branch cut crosses only occur if $-c_2\leq\operatorname{Im}(z)\leq c_2$ in both cases. This can be summarized by including additional terms:

\begin{equation}
    \begin{split}
    \overline{\alpha_s(z)} = \frac{-\kappa_\star i}{\pi}\Big[&(c-z)\log(c-z) - (\overline{c}-z)\log(\overline{c}-z)\\
    &+(-c-z)\log(-c-z) -(-\overline{c}-z)\log(-\overline{c}-z)\\
    &-2\pi i\cdot\left(c_1+\operatorname{Re}(z)\right)\cdot B_{(c_1,c_2)}(z)\\
    &- 2\pi i\cdot 2c_1\cdot B_{(\infty,c_2)}(z)\cdot H(\operatorname{Re}(z)-c_1)\Big]
    \end{split}
\end{equation} where $B_{(a,b)}$ is a two dimensional boxcar function \begin{align*}
    B_{(a,b)}(z) = \begin{cases}
                1,& -a \leq \operatorname{Re}(z) \leq a,\ -b \leq \operatorname{Im}(z) \leq b\\
                0,& \text{everywhere else}
            \end{cases}
\end{align*} and $H(x)$ is the Heaviside step function, \begin{align*}
    H(x) = \begin{cases}
                1,& x \geq 0\\
                0,& x < 0
            \end{cases}
\end{align*} Conjugating and factoring some minus signs to keep $-\kappa_\star$ apparent, we find

\begin{equation}
    \begin{split}
    \alpha_s(z) =&\begin{aligned}[t]
        \frac{-\kappa_\star i}{\pi}\Big[&(c-\overline{z})\log(c-\overline{z})-(\overline{c}-\overline{z})\log(\overline{c}-\overline{z})\\
        &+(-c-\overline{z})\log(-c-\overline{z})-(-\overline{c}-\overline{z})\log(-\overline{c}-\overline{z})\Big]
    \end{aligned}\\
    &-\kappa_\star\cdot (c + \overline{c} + z + \overline{z})\cdot B_{(c_1,c_2)}(z)\\
    &-\kappa_\star\cdot 2(c + \overline{c}) \cdot B_{(\infty,c_2)}(z)\cdot H(\operatorname{Re}(z)-c_1)
    \end{split}
\end{equation} Given that the region of interest in the image plane for IRS or IPM is always within the rectangle of the microlenses, this can be further simplified to \begin{equation}
    \begin{split}
    \alpha_s(z) =&\begin{aligned}[t]
        \frac{-\kappa_\star i}{\pi}\Big[&(c-\overline{z})\log(c-\overline{z})-(\overline{c}-\overline{z})\log(\overline{c}-\overline{z})\\
        &+(-c-\overline{z})\log(-c-\overline{z})-(-\overline{c}-\overline{z})\log(-\overline{c}-\overline{z})\Big]
    \end{aligned}\\
    &-\kappa_\star\cdot (c + \overline{c} + z + \overline{z})
    \end{split}
\end{equation} This is not the case when calculating the critical curves of the microlenses, which can extend outside the region of microlenses; see, e.g., \inprept{Weisenbach}.

\section{The fast multipole method}
\label{app:fmm}

For convenience, we give here the equations for the coefficients of the multipole and local expansions as derived by \cite{1987JCoPh..73..325G}, with a few comments for our implementation. 

\subsection{Creating the tree}

\begin{figure}
    \centering
    \includegraphics[width=\columnwidth]{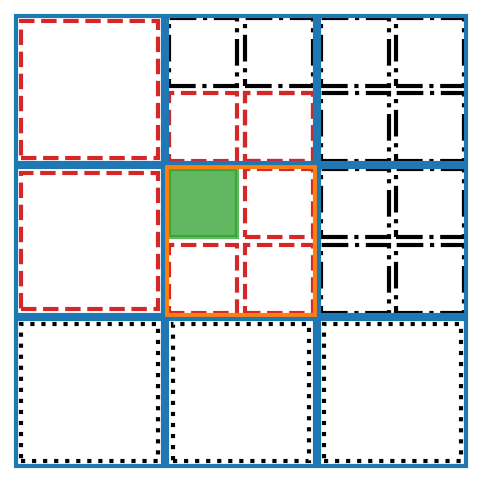}
    \caption{Visualization of some members of the tree at a given level (smaller squares) and one level above (larger squares). For a given node on the lower level (solid green, filled), its parent (central larger square, solid orange) has 8 neighbors (solid blue). Due to over-densities in the number of microlenses per unit area in different regions, only some of the parent's neighbor nodes are divided into smaller children. Regardless of size, some nodes from both levels share a side or corner with the given node and hence are neighbors to it (dashed red). The remaining nodes on the lower level (black, dot-dashed) are well separated from the given node, and their multipole coefficients can be transformed into Taylor series valid within the given node. The remaining nodes in the upper level (black, dotted) are not well-separated from the given node, and their multipole coefficients cannot be used to create a Taylor series within the given node. However, they contain a small number of microlenses (as they did not require further subdivision) which can instead be directly used to create a Taylor series.}
    \label{fig:tree_structure}
\end{figure}

When creating the tree we start with the root node, which is a square centered at the origin that is large enough to contain all the microlenses. The tree is then created by dividing each node into 4 children when necessary, until the maximum number of microlenses to be directly used for any node (i.e. the number of microlenses within the node and its neighbors) is no greater than some predefined limit. At every level in the tree, not every node will be split into children -- only those nodes which are over-dense or neighbor over-dense regions, in terms of the number of microlenses to use directly, will be further divided. An example of some members of the tree is shown in Figure~\ref{fig:tree_structure}.

\subsection{Multipole and local expansion coefficients}

At the lowest level in the tree, the potential due to the microlenses within a given node at a location $z$ far from the node is approximated by the multipole expansion \begin{equation}
    \psi(z) = a_0 \log z + \sum_{k=1}^{p}\frac{a_k}{z^k}
\end{equation}  truncated at some power $p$, where the multipole coefficients $a_k$ are directly found from  the $n_\star$\footnote{Here we use $n_\star$ to denote the number of microlenses within a node, as opposed to $N_\star$ for the entire region of microlenses.} microlenses of masses $m_i$ and locations $z_i$ as \begin{equation}
    \begin{split}
    a_0 &= \sum_{i=1}^{n_\star}m_i\\
    a_k &= \frac{1}{k}\sum_{i=1}^{n_\star}-m_i z_i^k, \quad k\geq 1
    \end{split}
\end{equation} The locations $z$ and $z_i$ are in units of the node half-length\footnote{We tend to use half-lengths, for symmetry purposes.}, and relative to the node center. The calculation of the multipole coefficients for each node can be done in parallel. 

The multipole coefficients of a parent node are found by adding together the shifted multipole coefficients of its 4 children. The shifted coefficients $b_l$ are \begin{equation}
    \begin{split}
    b_0 &= a_0\\
    b_l &= z_0^l\left[-\frac{a_0}{l}+\sum_{k=1}^{l}\frac{a_k}{z_0^k}\begin{pmatrix}l-1 \\ k-1\end{pmatrix}\right], \quad l\geq 1
    \end{split}
\end{equation} where $z_0$ is the center of the child node, relative to the center of the parent node, in units of the child node half-length. We precompute all of the binomial coefficients required for the maximum order $p$ as they will be reused many times, and use Horner's scheme to minimize the number of additions and multiplications necessary. Furthermore, in order to minimize potential losses from floating point precision, coefficients are always normalized to units of the node half-length. This means that, once a coefficient $b_l$ has been calculated in units of the child node half-length, it must be divided by $2^l$ since the parent node has twice the side length of its children. The shifted coefficients are again computed in parallel, and a single thread adds the coefficients of the children together to get the multipole coefficients of the parent. This process is continued for all nodes from the lowest level up to the root node; in addition, each time we move up a level in the tree, any node which did not have children has its multipole coefficients calculated as necessary. At this point, the multipole coefficients of every node are known, but all of the microlenses have only been used once within their lowest level node -- this is the power of the fast multipole method!

Next, the tree is traversed in the opposite direction. Inside a given node, the potential from distant nodes which are not neighbors is locally approximated by a Taylor series \begin{equation}
    \psi(z)=\sum_{l=0}^{p}c_l z^l
\end{equation} Each given node has a list of nodes in its `interaction list', which is essentially a list of the node's parent's neighbor's children which are not themselves neighbors to the given node; there are at most 27 nodes in this interaction list that are on the same level in the tree (i.e. of the same size), see \cite{1987JCoPh..73..325G} and Figure~\ref{fig:tree_structure} (black dot-dashed squares). For our adaptive tree, we have an additional interaction list which can contain at most 5 nodes from one level up in the tree (Figure~\ref{fig:tree_structure}, black dotted squares). This comes from the fact that not every node has children -- only those neighboring overdense regions require further subdivision; see Figure~\ref{fig:tree_structure}. 

The multipole coefficients of the nodes in the same level interaction list are converted into local coefficients $c_l$ for the given node as \begin{equation}
    \begin{split}
    c_0 &= b_0 \log(-z_0) + \sum_{i=1}^{p}\frac{b_k}{(-z_0)^k}\\
    c_l &= \frac{1}{z_0^l}\left[-\frac{b_0}{l} + \sum_{k=1}^{p}\frac{b_k}{(-z_0)^k}\begin{pmatrix}l + k -1 \\ k-1\end{pmatrix}\right], \quad l\geq 1
    \end{split}
\end{equation} where $z_0$ is the center of a node in the interaction list with respect to the given node. Local coefficients are again normalized to units of the node half-length. Since these coefficients come from nodes on the same level, this means that $c_l$ for $l\geq 1$ needs no additional changes, but we must further add $b_0\log L$ to $c_0$, where $L$ is the node half-length.

Nodes in the different level interaction list cannot have their multipole coefficients converted into local coefficients, as their node size is too large. Instead, the local coefficients within the given node are calculated directly from the microlenses in the distant node as \begin{equation}
    \begin{split}
    c_0 &= \sum_{i=1}^{n_\star}m_i \log(-z_i) + m_i\log{L}\\
    c_l &= \sum_{i=1}^{n_\star}\frac{-m_i}{l\cdot z_i^l}, \quad l\geq 1
    \end{split}
\end{equation} where the $z_i$ are in units of, and $L$ is the half-length of, the given node for which we are calculating the local coefficients. This is perhaps slightly computationally inefficient as we occasionally have to loop over microlenses again, but given the small number in each cell which must be processed this way, we find the potential memory savings outweighs the additional processing cost (especially when performing the computations on GPUs).

Finally, when we step down one level in the tree, the local coefficients of a parent node are shifted to create new local coefficients $d_l$ as \begin{equation}
    d_l = \frac{1}{(-z_0)^l}\sum_{k=l}^{p}c_k(-z_0)^k\begin{pmatrix}k \\ l\end{pmatrix}, \quad l\geq 0
\end{equation} which must be added onto the local coefficients of its children. Again, $z_0$ is the center of the child node, relative to the parent node, in units of the parent node half-length. The coefficients $d_l$ are also normalized to the child node half-length, which requires dividing them by $2^l$. 

\subsection{Determing the expansion order}

The expansion order $p$ determines the error introduced by cutting off the series expansions. \cite{1995PMPS.1934...389P} derive an estimate for the absolute error $|\epsilon|$ of the deflection angle calculated within a given node due to a single mass $m$ in a distant node on the same level, which we can write as \begin{equation}
    |\epsilon| \leq \frac{\theta_\star^2 m}{2L}\cdot 2\left(\frac{1}{2}\right)^{p} = \frac{\theta_\star^2 m}{L}\left(\frac{1}{2}\right)^p
\end{equation} where $L$ is the node half-length. This is not a perfect error estimate when considering the fact that we have many distant nodes which can each contain many masses; however, we might reasonably expect some of the errors from distant nodes on either side of a given node to cancel out due to symmetry. This means that, given some required error on the deflection angle $\epsilon$, we can reasonably take \begin{equation}
    p > \log_2\left(\frac{\theta_\star^2\frac{\langle m^2\rangle}{\langle m\rangle}}{L|\epsilon|}\right)
\end{equation} where the mass used takes into account the mass spectrum.

\section{Expansion near a macrocaustic}
\label{app:macrocaustic}

\begin{figure*}
    \centering
    \includegraphics[width=2\columnwidth]{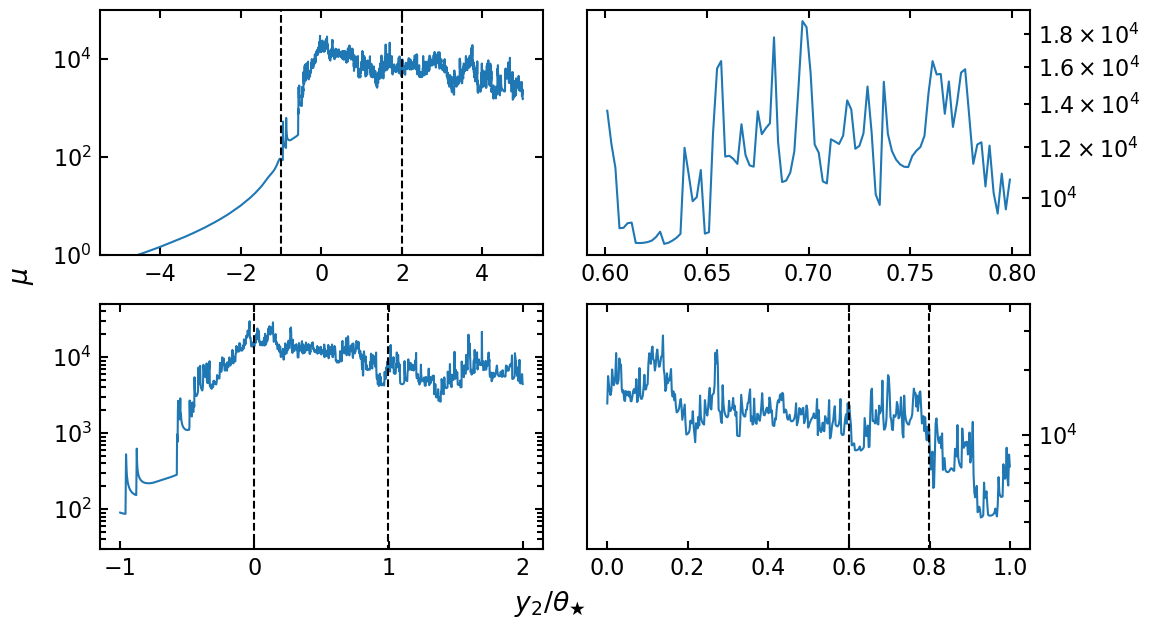}
    \caption{Microlensing lightcurve for a source moving perpendicular to the macrocaustic of Figure~\ref{fig:macro_ipm_maps} with $y_1=0$. Starting at the top left, successive zooms of various parts of the lightcurve (indicated by the vertical dashed lines) are shown in counterclockwise order. The pixel resolution here is insufficient to resolve caustic crossings at the finest scale, but still makes visible their vast abundance.}
    \label{fig:macro_ipm_lightcurves}
\end{figure*}

\begin{figure*}
    \centering
    \includegraphics[width=2\columnwidth]{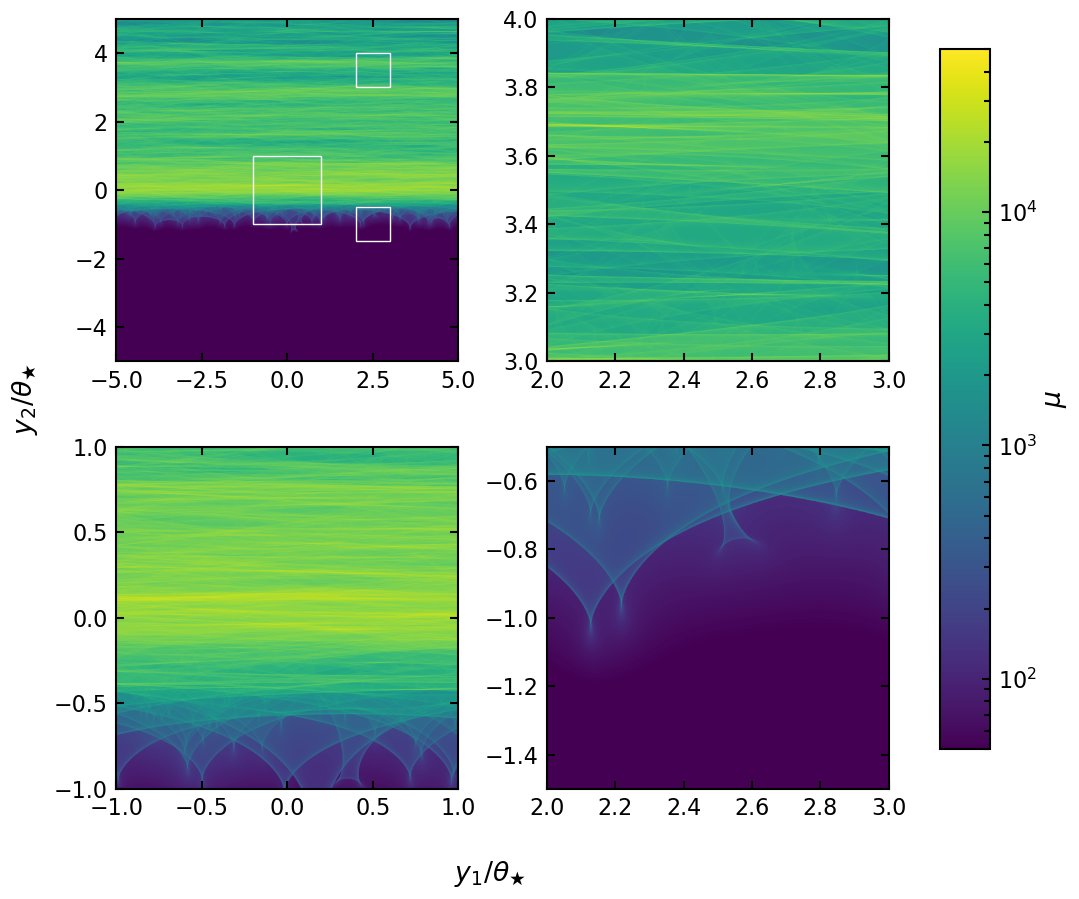}
    \caption{Microlensing magnification map in the vicinity of a macrocaustic located at $y_2=0$. Microlensing perturbs the macrocaustic into a network of microcaustics. See Figure~\ref{fig:macro_ipm_lightcurves} for an example lightcurve of a source moving perpendicular to the macrocaustic. The white squares in the top left subfigure indicate the regions whose zooms are shown in the top right, bottom right, and bottom left subfigures.}
    \label{fig:macro_ipm_maps}
\end{figure*}

Microlenses in the intracluster medium can disrupt the critical curve of galaxy clusters, creating a network of microcritical curves and turning the macrocaustic into a network of microcaustics \citep{2017ApJ...850...49V}. While Section~\ref{sec:extreme_mag_ex} touched on the case of such extreme magnification, it still only considered situations where gradients of the convergence and/or shear did not drastically vary over the image plane region under consideration. In situations near the critical curve where that is no longer true, the same general principles for simulating the effect of microlensing apply with a few modifications. 

A full discussion of simulations in this regime is outside the scope of this paper and better left to other works. The main point that differs from discussion up to now is how to properly determine the region(s) in the image plane within which to shoot rays; this depends on 1) the chosen order to which the potential is Taylor expanded, and 2) how one chooses to invert the non-linear lens equation to determine the boundary of the macroimages of the source plane region, either via algebraic or perturbative means. There are also minor caveats as to how one might wish to handle resolved or unresolved macroimages. 

A sample magnification map for the region around a macrocaustic is shown in Figure~\ref{fig:macro_ipm_maps}, with a lightcurve shown in Figure~\ref{fig:macro_ipm_lightcurves}. Creating the map in double precision took $\sim$1 hour due to the large extent of the image plane regions required. The location of the macrocaustic is easily visible, though closer examination shows how it is perturbed and composed of a multitude of microcaustics. Similar magnification maps can be seen in, e.g., \citet{1990PhDT.......180W}, \citet{2019A&A...625A..84D}, and \citet{2023ChA&A..47..570Y}.

\bsp	
\label{lastpage}
\end{document}